\documentclass[aps,showpacs,prb,twocolumn]{revtex4}%
\usepackage{amsfonts}
\usepackage{amsmath}
\usepackage{amssymb}
\usepackage{wasysym}
\usepackage{graphicx}%
\begin{document}
\title{Haldane's Instanton in 2D Heisenberg model revisited: along the Avenue of Topology}
\author{Ying Jiang}
\email{yjiang@shu.edu.cn} \affiliation{Department of Physics,
Shanghai University, Shanghai 200444, P.R. China} \affiliation{Key
Lab for Astrophysics, Shanghai 200234, P.R. China}
\author{Guo-Hong Yang}
\affiliation{Department of Physics, Shanghai University, Shanghai
200444,
 P.R. China} \affiliation{Key Lab for Astrophysics, Shanghai
200234, P.R. China}

\begin{abstract}
Deconfined quantum phase transition from N\'{e}el phase to Valence
bond crystal state in 2D Heisenberg model is under debate
nowadays. One crucial issue is the suppression of Haldane's
instanton on quantum critical point which drives the spinon
deconfined. In this paper, by making use of the $\phi$-mapping
topological current theory, we reexamine the Haldane's instanton
in an alternative way along the direction of topology. We find
that the the monopole events are space-time singularities of
N\'{e}el field $\vec{n}$, the corresponding topological charges
are the wrapping number of $\vec{n}$ around the singularities
which can be expressed in terms of the Hopf indices and Brouwer
degrees of $\phi$-mapping. The suppression of the monopole events
can only be guaranteed when the $\phi$-field possesses no zero
points. Moreover, the quadrapolarity of monopole events in the
Heisenberg model due to the Berry phase is also reproduced in this
topological argument.
\end{abstract}

 \maketitle

It is no doubt that Lev Landau was one of the most important
physicist on the Planet. His work on order parameters
\cite{landau-1}, as pointed by Hagen Kleinert
\cite{kleinert-landau}, was crucial for the development of the
modern theory of phase transition. Indeed, associated by the
powerful renormalization group theory, developed by Wilson, the
Landau-Ginzburg-Wilson (LGW) theory \cite{wilson-1} has become the
guiding principle behind the modern theory of critical phenomena.
Within this Landau's classic paradigm, the phase transitions are
described by fluctuations of physical order parameters which
reflect the symmetry breaking of the systems. This approach is
also adapted to investigate quantum critical phenomena as well and
provides the generally accepted framework for theoretical
descriptions of quantum phase transition
\cite{sachdev-quantum-phase-transition}.

However, in recent years, several possible exceptions beyond the
LGW paradigm, though under debate, have been discussed and have
been attracting lot of effort \cite{sachdev-nature}.

As is known, over the last two decades, a large amount of work has been
devoted to the study of low-dimensional quantum systems with frustrated
magnetic interactions. These systems, due to the quantum fluctuations and
reduced spatial dimensionality, have propensity to present variety of novel
states, such as the absence of magnetic order even at zero temperature, the
fractional elementary excitations, or the spontaneous dimerization of spins in
the valence bond crystal ground states \cite{misguich-1,bose-1}. The question
of spin-liquid states in two dimensions has been greatly motivated by the
high-$T_{c}$ superconductivity in the cuprates. Starting with Anderson's
proposal of the resonating valence bond theory of superconductivity
\cite{anderson-1}, a extensive amount of theoretical investigations are
centered around the notion that the superconducting state in these materials
is derived from a spin-liquid state by doping it with charge carriers. This
idea continues to inspire the search for new quantum spin models with novel
ground states on two-dimensional lattices and newer methods for studying these models.

In fact, prior to the motivations from the cuprates, low dimensional quantum
spin models, in their own right, have been attracting many people and efforts
for the reasons of explaining magnetic phenomena in real materials
\cite{kumar-1}. Purely theoretical motivations have led to interesting and
important developments in this field
\cite{lieb-1,affleck-1,haldane-1,hastings-1}. Among them, a lot of attention
has been focused particularly on the valence bond crystal (VBC) states, and
more precisely on the transition between these states and a magnetically
ordered phase \cite{read-1,read-2,nogueira-1}. In a VBC state, spins are
coupled in pairs forming singlet states, these pairwise singlets are called
valence bonds. These valence bonds themselves may be arranged in a periodic
pattern which breaks the translation symmetry. Such states then define an
order parameter quantifying the singlet long range order.

Since the ground state of the nearest neighbor Heisenberg antiferromagnet on
square lattice has the N\'{e}el order \cite{manousakis-1}, the VBC phase is
expected to emerge only when the ordered state is sufficiently frustrated by
adding suitable competing exchange interactions, i.e. the Hamiltonian under
investigation takes the form of%
\begin{equation}
H=J\sum_{\left\langle ij\right\rangle }\mathbf{S}_{i}\mathbf{\cdot S}%
_{j}+\cdots,
\end{equation}
where $\left\langle {}\right\rangle $ denotes the neareast neighbor pair, and
the ellipses stands for other competing interactions which leads to
frustration and may be tuned to drive quantum phase transitions.

The question on the nature of quantum phase transition that separates the VBC
state from a long-range ordered magnetic phase is very interesting and
important since it is crucial for understanding the high-temperature
superconductivity. In the Landau-Ginzburg approach, these two phases in the
ground state are independently characterized by the nonzero values of the
respective order parameters and the fluctuations thereof. Supposing a second
order transition within the Landau-Ginsburg paradigm \cite{landau-1}, the
order parameters of both phases should vanish precisely at the transition
point. Intuitively, it seems more probable that, without fine-tune, the two
order parameters will not vanish exactly at the same point, leading either to
a first order transition, or to two second order transitions separated by an
intermediate phase.

About five years ago, for a spin-1/2 system on square lattice, an
alternative yet interesting scenario, in which an ordered AF phase
undergoes a direct second order transition to VBC phase through a
common quantum critical point, has been proposed by Senthil and
co-workers \cite{senthil-1}. As discussed above, this direct
un-fine-tuned continuous transition between two-ordered phases,
which break different symmetries, could not be described in the
framework of Landau-Ginsburg theory \cite{senthil-1,senthil-2}.
Instead, in this scenario, however, such a quantum phase
transition is described by means of fractional degrees of freedom,
namely spinons. These spinons become deconfined at the quantum
critical point from which AF and the VBC phases are claimed to
emerge as the interaction parameters vary \cite{levin-1}. In this
exotic scenario of quantum phase transition, an essential point is
the behavior of Haldane's monopole events \cite{haldane-2} in
square lattice Heisenberg models. Senthil and coworkers argued
that, due to the Berry phase, these monopole events (or namely,
instantons) are suppressed exactly at quantum critical point,
causing the spinons deconfined which in turn leads to a continuum
quantum phase transition from long range order antiferromagnetic
phase to a VBC state \cite{senthil-1,senthil-2}.

The key feature of this non-Landau critical behavior is that at
the critical point the field theory in terms of fractionalized
objects with no obvious physical probe is a more appropriate
description. In spite of the difficulty of probing the
fractionalized excitations, the fractionalized nature of the
critical point leads to enormous anomalous dimension of the
physical order parameter that is distinct from the Wilson-Fisher
fixed-point or the mean-field result, which can be checked
experimentally.

However, recent renormalization group argument shows that the
above mentioned spinon deconfinement can only lead to a weak first
order phase transition \cite{flavio-2}. This viewpoint is
supported by Monte Carlo numerical simulation \cite{kuklov-2}. In
fact, doubts on the second order property of the deconfined
quantum phase transition never stopped immediately after it being
proposed \cite{kuklov-1,sirker-1,krueger-1}.

But, due to the lack of exact results, this issue keeps to be
controversial. For instance, recent numerical results suggest that
this transition may exist in a SU(2)
spin-1/2 model with both Heisenberg and ring exchange \cite{sandvik-1,melko-1}%
. The direct quantum phase transition between collinear N\'{e}el
phase and VBC states in two dimensional $J_{1}$-$J_{2}$ model has
been observed numerically by the use of exact diagonalization
\cite{gelle-1}. The quantum phase transition beyond the Landau's
paradigm in the spin-3/2 cold atom systems has also been
investigated with the help of Sp(4) spin Heisenberg models on the
triangular and square lattices \cite{qi-1}.

Nevertheless, as pointed above, the crucial feature of the quantum
phase transition between the magnetic ordered phase and VBS is the
deconfinement of spinons which relies on the suppression of
Haldane's monopole events at the quantum phase transition point.
Hence, it is important to revisit the Haldane's monopole events
from a more fundamental viewpoint, i.e. along the avenue of
topology.

As is known, in the frame of field theory, the long-wavelength
fluctuations around the direction of the N\'{e}el order parameter
of the two-dimensional quantum Heisenberg spin system can be
descriped by the $O\left(  3\right)  $ non-linear $\sigma$-model.
By virtue of spin coherent state path integral, the
corresponding action is expressed as%
\begin{equation}
\mathcal{A}=\frac{1}{2}\int d\tau\int d^{2}r\left[  \frac{1}{c^{2}}\left(
\frac{\partial\vec{n}}{\partial\tau}\right)  ^{2}+\left(  \nabla_{r}\vec
{n}\right)  ^{2}\right]  +S_{B}, \label{action}%
\end{equation}
where $S_{B}$ is the so-called Berry phase
\cite{sachdev-quantum-phase-transition} which strongly affects the behavior of
the integer and half integer spin system \cite{haldane-1,haldane-2}. The Berry
phase is unimportant in the long-wave length properties of N\'{e}el state
\cite{chakravarty-1}, while is very crucial in describing the quantum
paramagnetic phase \cite{read-2}. Thus, the $O(3)$ non-linear $\sigma$-model
associated with the Berry phase is sufficient to be used to investigate the
N\'{e}el state and VBC state.

As is known, a N\'{e}el configuration $\vec{n}\left(  \vec{r}\right)  $ of the
system can be characterized by the corresponding Pontryagin index%
\begin{equation}
Q=\frac{1}{4\pi}\int d^{2}r\vec{n}\cdot\left(  \partial_{x}\vec{n}%
\times\partial_{y}\vec{n}\right)  \label{pontryagin}%
\end{equation}
which can be written in a more symmetric way%
\begin{equation}
Q=\frac{1}{8\pi}\int d^{2}r\epsilon^{0\mu\nu}\epsilon^{abc}n^{a}\partial_{\mu
}n^{b}\partial_{\nu}n^{c}, \label{topological-current}%
\end{equation}
here $a,b,c$ denote the indices in spin space, $\mu$ and $\nu$ stand for the
lattice spacial indices while $0$ for the time parameter $\tau$. This quantity
is interpreted as the total topological charge of a type of topological
defects in this system, skyrmions. The topological charge totally reflects the
topological structure of the N\'{e}el configuration of the non-linear $\sigma$-system.

The Pontryagin index is a topological invariant for a continuous manifold, in
other words, in the $O\left(  3\right)  $ non-linear $\sigma$-model, providing
that $\vec{n}\left(  \vec{r}\right)  $ is always well defined in the whole
space, it is easy to verify that the total topological charge doesn't change
as the field evolves with time, i.e. the during any time interval, we have%
\begin{equation}
\Delta_{\tau}Q=0. \label{conserve}%
\end{equation}

However, when we look back to our original system, we recognize that in the
original Heisenberg model, spins sit only on lattice sites, when considering
the long-wavelength fluctuations and taking continuum limit, singularities of
configurations $\vec{n}\left(  \vec{r}\right)  $ at places away from the
lattice sites may be allowed, which will violate the topological charge
conservation of skyrmions in the system, and the so-called monopole events or
instantons emerge.

In order to investigate topological properties of the system properly in the
framework of non-linear $\sigma$ model, the singularities mentioned above
should be taken into account, hence the condition of well defined $\vec
{n}\left(  \vec{r}\right)  $ is relaxed. In this case, generally, the N\'{e}el
configuration $\vec{n}\left(  \vec{r}\right)  $ would be expressed as the norm
of a specific underlying field $\vec{\phi}\left(  \vec{r}\right)  $,%
\begin{equation}
n^{a}=\frac{\phi^{a}}{\left\Vert \phi\right\Vert }. \label{phi}%
\end{equation}
In this way, the singularities have been properly encrypted in, and the
singular points of $\vec{n}$ field correspond to zero points of $\vec{\phi}$
field. Then, the topological charge density of skyrmions in Eq.
(\ref{topological-current}) is rewritten in terms of $\vec{\phi}$ as%
\begin{equation}
q=\frac{1}{8\pi}\epsilon^{0\mu\nu}\epsilon^{abc}n^{a}\partial_{\mu}%
n^{b}\partial_{\nu}n^{c}=\frac{1}{8\pi}\epsilon^{0\mu\nu}\epsilon^{abc}%
\frac{\phi^{a}}{\left\Vert \phi\right\Vert ^{3}}\partial_{\mu}\phi^{b}%
\partial_{\nu}\phi^{c}. \label{topological-charge-density}%
\end{equation}
By making use of the topological current method
\cite{jiang-1,jiang-2,jiang-3} and Laplacian Green function
relation, straightforward calculation shows that
\begin{equation}
\partial_{\tau}q=D\left(  \frac{\phi}{z}\right)  \delta\left(  \vec{\phi
}\right)  , \label{delta-structure}%
\end{equation}
where $D\left(  \phi/z\right)  $ is the Jacobian of the
$\phi$-mapping, here $z$ denotes $\left(  x,y,\tau\right)  $. The
equation Eq. (\ref{delta-structure}) clearly shows that, when
$\vec{\phi}$ field possesses no zero point, i.e. when $\vec{n}$ is
well defined in the whole space-time, the change rate of the
skyrmion charge is zero, as what we expect; while when
$\vec{\phi}$ field does possess zero point, the change rate of the
skyrmion charge is no longer zero, it entirely depends on the
mapping properties of the $\vec{\phi}$-field at the zero point, in
other words, depends on the topological property of the
$\vec{\phi}$-field. These singularities of $\vec{n}$ are just the
so-called monopole events or instantons. This clearly shows that
the conservation of skyrmions thence the deconfinement of spinons
are closely related to the absence of zero points of $\vec{\phi}$
field.

To determine the relation between the change rate of the skyrmion number and
the topological index of the $\phi$-mapping is definitely an interesting and
constructive topic. Actually, the change rate of the skyrmion number is just
the space-time charge density of corresponding monopole events. In order to
investigate the change rate of the skyrmion number, let us suppose that the
$i$-th instanton emerges at the space-time point $z_{i}=\left(  \tau_{i}%
,x_{i},y_{i}\right)  $, i.e. at these points, $\vec{\phi}\left(  z_{i}\right)
=0$, then the $\delta$-function $\delta\left(  \vec{\phi}\right)  $ can be
decomposed as%
\begin{equation}
\delta\left(  \vec{\phi}\right)  =\sum_{i=1}^{l}c_{i}\delta\left(
z-z_{i}\right)  ,\ \ \ c_{i}=\frac{\beta_{i}}{\left\vert D\left(  \frac{\phi
}{z}\right)  _{z_{i}}\right\vert } \label{delta}%
\end{equation}
where $\beta_{i}$ is the Hopf index of the $\phi$-mapping around the zero
point $z_{i}$. When substituting this relation into Eq. (\ref{delta-structure}%
), we find that
\begin{equation}
\partial_{\tau}q=\sum_{i}\beta_{i}\eta_{i}\delta\left(  z-z_{i}\right)  ,
\label{delta-rate}%
\end{equation}
where $\eta_{i}=\mathrm{sign}\left[  D\left(  \frac{\phi}{z}\right)  _{z_{i}%
}\right]  $ is the so-called the Brouwer degree of the $\phi$-mapping around
this zero point. $\beta_{i}\eta_{i}$ is interpreted as the topological charge
of the $i$-th monopole event located at space-time point $z_{i}$. Hence, we
can easily get that, in any finite time interval, the total change of the
skyrmion number is the sum of topological charges of all monopole events
emerging in that time interval, i.e.%
\begin{equation}
\Delta Q=\int d\tau\int d^{2}r\ \partial_{\tau}q=\sum_{i}\beta_{i}\eta_{i}.
\label{change-rate}%
\end{equation}
In fact, this is just the wrapping number of $\vec{n}$, due to the second
homotopy \cite{allen-1} of sphere $\pi_{2}\left(  S^{2}\right)  =Z$, it is
integer. The above equation shows that $Q$ is nonconserved, and monopole
events (or instantons) appear to change the skyrmion number $Q$ by integer value.

Now let us concentrate on the influence of the Berry phase to the behavior of
the spin system with integer and half-integer spins. The Berry phase has the
relation with $\vec{n}$ through $Q_{xt}$, which is analogous to the skyrmion
number $Q$ but in the two-dimensional $x$-$t$ space-time \cite{haldane-2}, as
is shown
\begin{equation}
S_{B}=2\pi S\sum_{n}\left(  -1\right)  ^{n}Q_{xt}\left(  y_{n}\right)  .
\label{berry-phase}%
\end{equation}
When taking continuum limit, we can immediately get
\begin{equation}
S_{B}=\frac{2\pi S}{2}\Delta Q. \label{quantization}%
\end{equation}
Associated with the single-valuedness of wavefunction of the
system, i.e., the Berry phase should be integer multiple of $2\pi$
to ensure the absence of destructive interference of monopole
events, this expression indicates clearly that
for spin $1/2$ system,%
\begin{equation}
\Delta Q=4m,\ \ \ m\in Z, \label{quadrapole}%
\end{equation}
i.e. the monopole events, when exist, should be quadrapoled.

The above simple argument shows a good agreement with Haldane's
result on square lattice Heisenberg model \cite{haldane-2} and
other previous subtle analysis \cite{read-1,read-2} , but from a
different point of view. Moreover, as pointed out by Senthil and
his collaborators \cite{senthil-1}, the compactness of the system
is violated by the destructive interference between the Berry
phase and monopole events, leading the skyrmion number
asymptotically conserved exactly on the quantum critical point,
and the quantum phase transition between N\'{e}el state and VBS
state occurs mediated by the deconfinement of the emergent spinon
degree of freedom. This scenario reflects the non-compactness of
the system at the QCP. From our calculation, this is due to the
absence of zero points of $\vec{\phi}$ field. To determine the
condition under which the zero points of $\vec{\phi}$ field are
suppressed may provide an alternative possibility to look insight
into the intrinsic properties of the deconfinement quantum
critical phenomena.

\section*{Acknowledgement}

Work supported by Shanghai Leading Academic Discipline Project(
project Number: S30105), and by NSFC under Grant No. 10845002.
Financial support from the Science \& Technology Committee of
Shanghai Municipality under Grant Nos. 08dj1400202, 09PJ1404700
(Y.J.) and 08JC14097 (G.-H.Y) is also acknowledged.

\end{document}